\documentclass{article}

\PassOptionsToPackage{numbers, compress}{natbib}
\usepackage[final]{neurips_2023}

\usepackage[utf8]{inputenc} % allow utf-8 input
\usepackage[T1]{fontenc}    % use 8-bit T1 fonts
\usepackage{hyperref}       % hyperlinks
\usepackage{url}            % simple URL typesetting
\usepackage{booktabs}       % professional-quality tables
\usepackage{amsfonts}       % blackboard math symbols
\usepackage{nicefrac}       % compact symbols for 1/2, etc.
\usepackage{microtype}      % microtypography
\usepackage{xcolor}         % colors
\usepackage{graphicx}
\newcommand{\MYhref}[3][blue]{\href{#2}{\color{#1}{#3}}}%

\title{Real-time Animation Generation and Control on Rigged Models via Large Language Models}

\author{%
 Han Huang \\
  Rensselaer Polytechnic Institute \& Microsoft \\
  \And
  Fernanda De La Torre \\
  MIT \& Microsoft\\
  \And
  Cathy Mengying Fang\\
  MIT Media Lab \& Microsoft \\
  \And
  Andrzej Banburski-Fahey \\
  Microsoft \\
  \And
  Judith Amores \\
  Microsoft \\
  \And
  Jaron Lanier \\
  Microsoft \\
}

\begin{document}

\maketitle

\begin{abstract}

We introduce a novel method for real-time animation control and generation on rigged models using natural language input. First, we  embed a large language model (LLM) in Unity to output structured texts that can be parsed into diverse and realistic animations. Second, we illustrate LLM's potential to enable flexible state transition between existing animations. We showcase the robustness of our approach through qualitative results on various rigged models and motions.
\end{abstract}

\section{Introduction}

While recent advances in deep learning have revolutionized many aspects of computer graphics \cite{jun2023shap, li20223ddesigner, lin2023magic3d, singer2023text, text2room, hong20233d, sra2017oasis, gao2022nerf, poole2022dreamfusion}, few works have explored direct actuation of rigged 3D models. In this context, we present an innovative framework that leverages large language models (LLM) to enable real-time animation generation and control.

First, we generate novel animations on a given model using only natural language descriptions. Our method outputs structured strings encoding positional and rotational time series for each joint, which are parsed to produce animations on the rigged object. We showcase the generated animations on hierarchically distinct models with a variety of motions to underscore the robustness of our approach. 

Second, we integrate a LLM with Unity  \cite{unity} to program animation transition on humanoid characters via the generation and execution of appropriate Unity C\# scripts. Our approach is characterized by its flexibility, allowing for the seamless integration of pre-existing animations with custom game logic.

\section{Methodology and Results}

\paragraph{Animation Generation}

\begin{figure}
    \centering
    \includegraphics[width=0.56\linewidth]{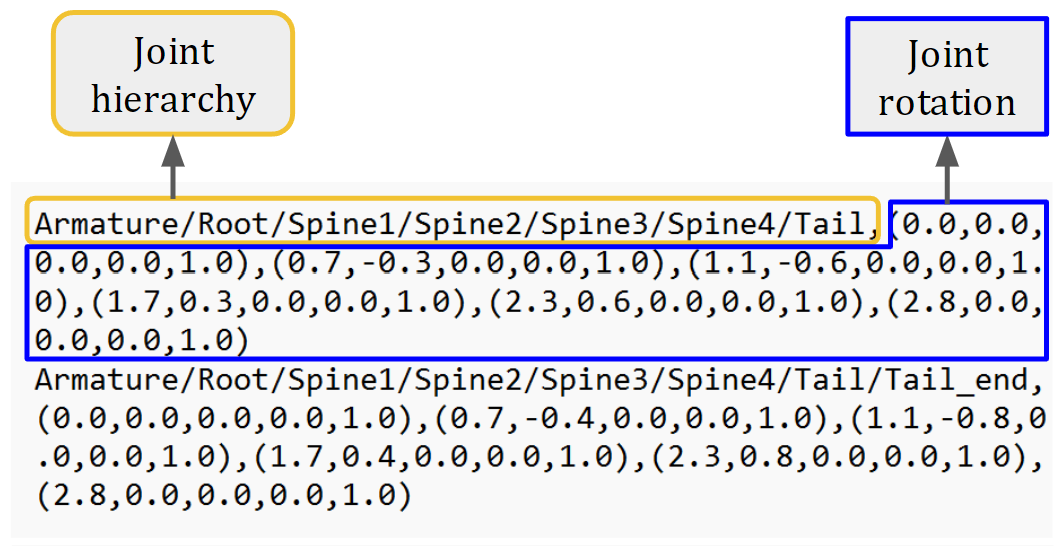}
    \includegraphics[width=0.15\linewidth]{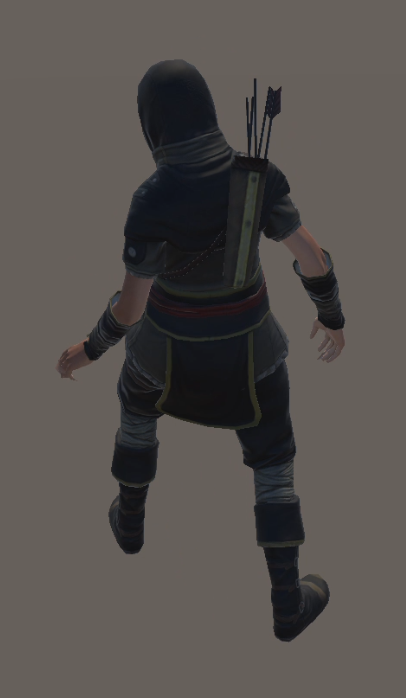}
    \includegraphics[width=0.126\linewidth]{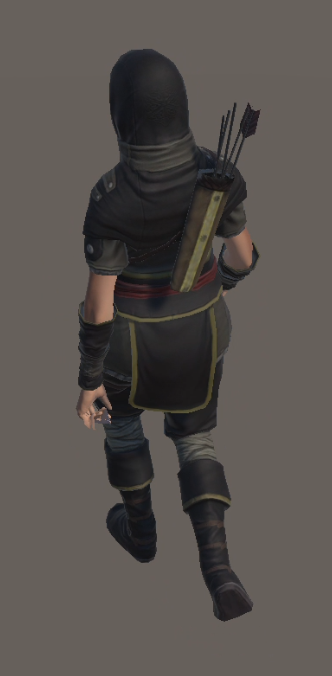}
    \includegraphics[width=0.141\linewidth]{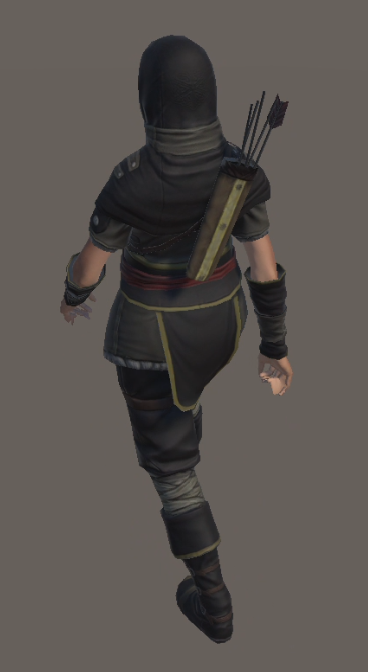}
    \caption{Left: animation string generated with the prompt: "make the whale flapping its tail" on a whale model. Each vector is a quarternion followed by a time stamp: $(q_0, q_1, q_2, q_3, t)$; Right: Animation control on a character with existing animations. Prompt: "Make the 'archer' move around randomly and pauses once in a while with correct animations."}
    \label{fig:anim_control_and_txt}
    \vspace{-4mm}
\end{figure}

Formally, we abstract a rigged 3D model as a tree $T = (V,E)$ encoding its joint hierarchy. An animation on the model is then a set of motion time series $(p^v_i, q^v_i)$ associated with each joint $v\in V$, where $p^v_i\in \mathbb{R}^3, q^v_i\in \mathbb{R}^4$ are the joint position and rotation (as a quaternion) at time $t_i\in [0,t_{end}]$, respectively. To generate animations on a rigged object, we utilize a LLM to output structured texts containing appropriate joint hierarchies and motions according to the user prompt. We show an example in Figure \ref{fig:anim_control_and_txt} and remark that numerical values are also treated as text tokens during generation. To overcome token size limits, we compress the animation strings to only output motion values at \textit{keyframes} and truncate all floating-point numbers to a single significant figure.

\begin{figure}
    \includegraphics[width=1\linewidth]{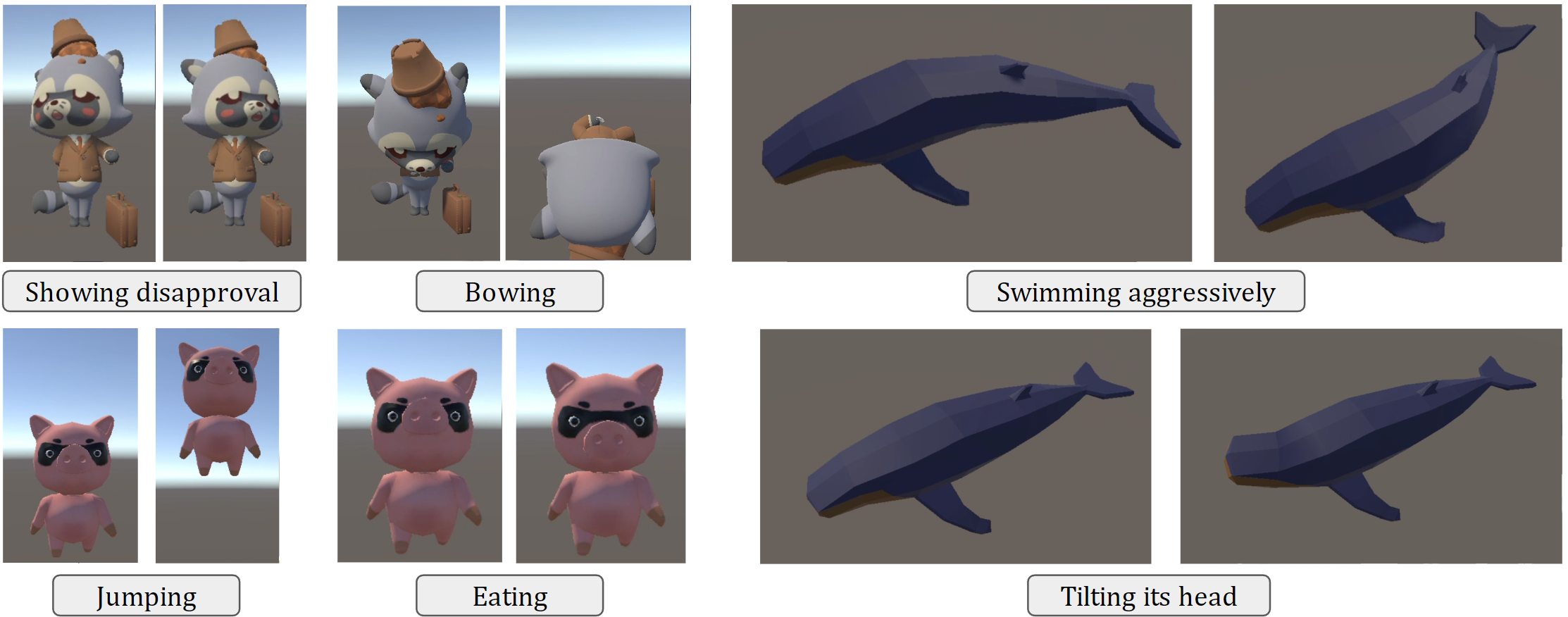}
    \vspace{-5mm}
    \caption{One-shot animation generation on rigged models. Text bubbles contain the prompts used.}
    \label{fig:few_shot_examples}
    % \vspace{-2mm}
\end{figure}

We illustrate novel animations generated for a whale, a pig, and a raccoon in Figure \ref{fig:few_shot_examples}. Here, each model comes with an existing animation, which we use for in-context learning\cite{GPT-3}. Overall, the generated motions are visually realistic and can be produced within a matter of seconds. Our framework also exhibits a high degree of semantic comprehension on joint hierarchies by producing motions on the appropriate bones. For instance, when prompted with "tilting its head" for a whale, our model translates this into motion within the head joint while keeping the main body still. Remarkably, the animated models are rich in structural diversity, which underscores the robustness of our framework in accommodating distinct anatomies and motion patterns. Our result suggests the LLM has successfully leveraged certain physical priors acquired during its training process.

\paragraph{Animation Control}
Next, we showcase LLM's ability to perform animation control, which seeks to integrate existing animation clips into the game logic. For example, a player-controlled character should enter the jumping animation when the "space" key is pressed. In this work, we follow \cite{prompt_based_creation_VR, roberts2022surreal, LLMR} to generate Unity C\# scripts with GPT-4 \cite{GPT-4}, then execute the composed scripts with the Roslyn run-time compiler to enable animation state transitions \cite{roslyn}. Figure \ref{fig:anim_control_and_txt} shows an archer model downloaded from Mixamo \cite{mixamo} programmed to transition from being idle to walking per user prompt. We remark that the generated control flow is flexible and can be retargeted to any humanoid characters via Unity's Avatar system \cite{unity}. 

\section{Discussion}

We encourage the reader to inspect videos for the generated animations on our \MYhref{https://github.com/Whalefishin/LLM_animation}{Github repository}. In addition, we provide more details on the results shown in Figures  \ref{fig:few_shot_examples} and \ref{fig:anim_control_and_txt} in the supplementary material.

To the best of our knowledge, our approach marks the pioneering method for actuating 3D models based on natural language descriptions. By embedding our framework in Unity, its output keyframes can be readily adjusted in the Unity editor window, allowing human animators to enhance and refine the results as needed. Hence, a potential application of our model is to produce initial drafts for digital artists similar to frameworks such as Midjourney and Stable Diffusion\cite{midjourney,stable_diff}.

\bibliographystyle{abbrv}
\bibliography{bb}

%%%%%%%%%%%%%%%%%%%%%%%%%%%%%%%%%%%%%%%%%%%%%%%%%%%%%%%%%%%%

\end{document}

\maketitle

\section{Animation Generation}

% In this section, we provide further details and insights into our animation generation approach. To view the generated animations in their entirety, please refer to the \verb|videos| folder, where all recorded clips are stored.

\subsection{Motion on a Rigged Model}

To harness LLM's outstanding capability in generalization and instruction-following, we introduce a text-based representation for motions on a joint hierarchy. To simplify the task, we consider the subset of motions that consist of joint rotation and root motion, which is translation on the entire rigged model. In particular, we disregard positional movement on the joints as they are unusual for typical objects and motions. However, our framework can be readily extended to scenarios with joint translations, which may be relevant for non-rigid objects such as an octopus.

\subsection{LLM Metaprompt}

Our approach leverages a LLM to generate structured strings representing joint hierarchies and motions, which is significantly different from the output distribution for a typical conversational agent. To achieve this, we utilize a metaprompt shown in \ref{anim-generator-metaprompt} to specify the LLM's specialized role and ensure syntactic validity in the generated samples. 

\begin{lstlisting}[style=mystyle, caption={Metaprompt for animation generation},label=anim-generator-metaprompt, frame=single]
You're an animator who will be provided the joints on a rigged 3D model, and you have to rotate them to produce the requested animation. The joints will be stored as a JSON string that outlines the object hierarchy. 
You will be given a few examples for animations on the object you need to animate, and you have to come up with new animations on the same object.

The object you will animate is a [OBJECT_NAME].

Object JSON: [OBJECT_JSON]

# Animation string
- The first line of the string specifies the root motion for the animation. It starts with the joint name for the armature root, followed by vectors in the format of [t,x,y,z] specifying a key frame. "t" is the time stamp for the key, and "x", "y", "z" give the x,y,z components for the position of the object root. For example, Armature,[0.0,0.0,0.0,5.9],[1.3,1.0, 2.4, 5.9] means that the root "Armature" has position [0.0,0.0,5.9] at time 0.0, and position [1.0, 2.4, 5.9] at time 1.3. 
- After the first line, each subsequent line of the string starts with the joint name, then each vector in the format (t,x,y,z,w) specify a key frame for the animation. "t" is the time stamp for the key, and "x", "y", "z", "w" give the x,y,z,w components for the rotation quaternion, respectively. For example, Armature/Root/Head,(0.0,0.7,0.0,0.0,0.7),(1.3,0.6,0.0,0.0,0.8) means that the joint "Armature/Root/Head" has rotation (0.7,0.0,0.0,0.7) at time 0.0, and rotation (0.6,0.0,0.0,0.8) at time 1.3. 

[ANIMATION_NAME]
[ANIMATION_STRING]
\end{lstlisting}
\lstset{style=mystyle}

There are four placeholders in the metaprompt. First, the model needs to know about the existing joints on the given object in order to animate it. This information is provided via \verb|OBJECT_NAME| and \verb|OBJECT_JSON|. Second,  \verb|ANIMATION_NAME| and \verb|ANIMATION_STRING| serves as in-context demonstrations. Note that these placeholders can be replaced for concrete examples with straightforward C\# scripts that parse the object hierarchy and animation clips into strings. A concrete example displayed in \ref{anim-generator-metaprompt-concrete} is used for animating the raccoon shown in the main text. \\

\begin{lstlisting}[style=mystyle, caption={Metaprompt for animation generation on a raccoon},label=anim-generator-metaprompt-concrete, frame=single]
You're an animator who will be provided the joints on a rigged 3D model, and you have to rotate them to produce the requested animation. The joints will be stored as a JSON string that outlines the object hierarchy. 
You will be given a few examples for animations on the object you need to animate, and you have to come up with new animations on the same object.


The object you will animate is a **raccoon**.

Object JSON:
name:metarig,position:(0.00,0.00,0.00),rotation:(-0.7,0.0,0.0,0.7),children:[name:spine,position:(0.00,0.00,0.00),rotation:(0.7,0.0,0.0,0.7),children:[name:pelvis.L,position:(0.00,0.00,0.00),rotation:(-0.2,0.6,0.7,0.4),name:pelvis.R,position:(0.00,0.00,0.00),rotation:(0.2,0.6,0.7,-0.4),name:spine.001,position:(0.00,0.00,0.00),rotation:(0.0,0.0,0.0,1.0),children:[name:spine.002,position:(0.00,0.00,0.00),rotation:(0.0,0.0,0.0,1.0),children:[name:spine.003,position:(0.00,0.00,0.00),rotation:(-0.1,0.0,0.0,1.0),children:[name:breast.L,position:(0.00,0.00,0.00),rotation:(0.0,0.8,0.6,0.0),name:breast.R,position:(0.00,0.00,0.00),rotation:(0.0,0.8,0.6,0.0),name:shoulder.L,position:(0.00,0.00,0.00),rotation:(-0.7,0.2,0.4,0.6),children:[name:upper_arm.L,position:(0.00,0.00,0.00),rotation:(0.2,-0.7,0.4,0.5),children:[name:forearm.L,position:(0.00,0.00,0.00),rotation:(0.5,0.0,0.0,0.8),children:[name:hand.L,position:(0.00,0.00,0.00),rotation:(0.1,0.0,-0.1,1.0)]]],name:shoulder.R,position:(0.00,0.00,0.00),rotation:(-0.7,-0.2,-0.4,0.6),children:[name:upper_arm.R,position:(0.00,0.00,0.00),rotation:(-0.1,0.9,-0.3,0.4),children:[name:forearm.R,position:(0.00,0.00,0.00),rotation:(-0.1,0.2,-0.6,0.8),children:[name:hand.R,position:(0.00,0.00,0.00),rotation:(0.1,0.1,-0.2,1.0),children:[name:hand.R.001,position:(0.00,0.00,0.00),rotation:(0.3,0.0,0.7,0.6),children:[name:Spork_low,position:(0.00,0.00,0.00),rotation:(0.0,0.7,-0.7,0.1)]]]]],name:spine.006,position:(0.00,0.00,0.00),rotation:(-0.1,0.0,0.0,1.0),children:[name:ear.L,position:(0.00,0.01,0.00),rotation:(-0.1,-0.1,0.2,1.0),name:ear.R,position:(0.00,0.01,0.00),rotation:(-0.1,0.1,-0.5,0.9)]]]],name:tail,position:(0.00,0.00,0.00),rotation:(0.8,0.4,-0.1,-0.3),children:[name:tail.001,position:(0.00,0.00,0.00),rotation:(-0.2,0.1,-0.2,1.0),children:[name:tail.002,position:(0.00,0.00,0.00),rotation:(0.0,0.5,-0.5,0.7),children:[name:tail.003,position:(0.00,0.00,0.00),rotation:(-0.5,0.0,0.0,0.8)]]],name:thigh.L,position:(0.00,0.00,0.00),rotation:(1.0,0.1,-0.3,0.0),children:[name:shin.L,position:(0.00,0.00,0.00),rotation:(0.2,0.3,-0.1,0.9),children:[name:foot.L,position:(0.00,0.00,0.00),rotation:(-0.5,0.1,0.2,0.8),children:[name:heel.02.L,position:(0.00,0.00,0.00),rotation:(-0.6,0.6,-0.2,-0.5),name:toe.L,position:(0.00,0.00,0.00),rotation:(-0.3,0.8,-0.4,-0.3)]]],name:thigh.R,position:(0.00,0.00,0.00),rotation:(1.0,-0.1,0.3,0.0),children:[name:shin.R,position:(0.00,0.00,0.00),rotation:(0.2,-0.3,0.1,0.9),children:[name:foot.R,position:(0.00,0.00,0.00),rotation:(-0.5,-0.1,-0.2,0.8),children:[name:heel.02.R,position:(0.00,0.00,0.00),rotation:(0.6,0.6,-0.2,0.5),name:toe.R,position:(0.00,0.00,0.00),rotation:(0.3,0.8,-0.4,0.3)]]]]]

# Animation string
- The first line of the string specifies the root motion for the animation. It starts with the joint name for the armature root, followed by vectors in the format of [t,x,y,z] specifying a key frame. "t" is the time stamp for the key, and "x", "y", "z" give the x,y,z components for the position of the object root. For example, Armature,[0.0,0.0,0.0,5.9],[1.3,1.0, 2.4, 5.9] means that the root "Armature" has position [0.0,0.0,5.9] at time 0.0, and position [1.0, 2.4, 5.9] at time 1.3. 
- After the first line, each subsequent line of the string starts with the joint name, then each vector in the format (t,x,y,z,w) specify a key frame for the animation. "t" is the time stamp for the key, and "x", "y", "z", "w" give the x,y,z,w components for the rotation quaternion, respectively. For example, Armature/Root/Head,(0.0,0.7,0.0,0.0,0.7),(1.3,0.6,0.0,0.0,0.8) means that the joint "Armature/Root/Head" has rotation (0.7,0.0,0.0,0.7) at time 0.0, and rotation (0.6,0.0,0.0,0.8) at time 1.3. 


idle while moving head up and down, animation: 

metarig,[0.00,0.00,0.00,0.00],[1.56,0.00,0.00,0.00]
metarig/spine/spine.001/spine.002/spine.003/spine.006,(0.0,-0.1,0.0,0.0,1.0),(0.5,-0.1,0.0,0.0,1.0),(0.8,-0.1,0.0,0.0,1.0),(1.2,0.0,0.0,0.0,1.0),(1.6,-0.1,0.0,0.0,1.0)
metarig/spine/spine.001/spine.002/spine.003/spine.006/ear.L,(0.0,-0.1,-0.1,0.1,1.0),(0.2,-0.1,0.0,0.0,1.0),(0.4,0.0,-0.1,0.1,1.0),(0.6,0.0,-0.1,0.2,1.0),(1.0,-0.1,-0.1,0.2,1.0),(1.2,0.0,-0.1,0.2,1.0),(1.5,-0.1,-0.1,0.1,1.0),(1.6,-0.1,-0.1,0.1,1.0)
metarig/spine/spine.001/spine.002/spine.003/spine.006/ear.R,(0.0,-0.1,0.0,-0.3,1.0),(0.2,-0.1,0.0,-0.3,0.9),(0.4,0.0,0.1,-0.2,1.0),(0.6,0.0,0.0,-0.2,1.0),(1.0,-0.1,0.0,-0.1,1.0),(1.2,0.0,0.0,-0.2,1.0),(1.5,-0.1,0.1,-0.2,1.0),(1.6,-0.1,0.0,-0.3,1.0)
metarig/spine/spine.001/spine.002/spine.003/shoulder.L/upper_arm.L,(0.0,-0.1,0.7,-0.5,-0.5),(0.5,-0.1,0.8,-0.4,-0.4),(0.9,-0.1,0.9,-0.4,-0.4),(1.3,-0.1,0.7,-0.5,-0.5),(1.6,-0.1,0.7,-0.5,-0.5)
metarig/spine/spine.001/spine.002/spine.003/shoulder.L/upper_arm.L/forearm.L,(0.0,0.1,-0.1,0.0,1.0),(0.4,0.1,-0.1,0.0,1.0),(0.8,-0.1,-0.2,0.0,1.0),(1.2,0.0,-0.2,0.0,1.0),(1.6,0.1,-0.1,0.0,1.0)
metarig/spine/spine.001/spine.002/spine.003/shoulder.L/upper_arm.L/forearm.L/hand.L,(0.0,0.1,0.0,-0.1,1.0),(0.5,0.1,0.0,0.0,1.0),(1.0,0.1,-0.1,0.1,1.0),(1.5,0.1,0.0,0.0,1.0),(1.6,0.1,0.0,-0.1,1.0)
metarig/spine/tail,(0.0,-0.8,0.0,0.0,0.6),(0.2,-0.8,0.0,0.0,0.6),(0.4,-0.9,-0.2,-0.2,0.5),(0.6,-0.9,-0.2,-0.2,0.4),(1.1,-0.8,-0.1,-0.1,0.5),(1.6,-0.8,0.0,0.0,0.6)
metarig/spine/tail/tail.001,(0.0,0.1,0.0,0.1,1.0),(0.4,0.1,0.0,0.0,1.0),(0.8,0.0,0.1,-0.2,1.0),(1.2,0.1,0.0,0.0,1.0),(1.6,0.1,0.0,0.1,1.0)
metarig/spine/tail/tail.001/tail.002,(0.0,0.1,0.0,0.2,1.0),(0.4,0.2,0.0,0.3,0.9),(0.6,0.1,0.0,0.1,1.0),(0.7,0.1,0.0,-0.1,1.0),(1.0,0.0,0.0,-0.4,0.9),(1.3,0.1,0.0,-0.1,1.0),(1.5,0.1,0.0,0.1,1.0),(1.6,0.1,0.0,0.2,1.0)
metarig/spine/tail/tail.001/tail.002/tail.003,(0.0,0.0,0.1,0.2,1.0),(0.4,0.0,0.0,0.5,0.9),(0.6,0.0,0.0,0.5,0.9),(0.7,0.0,0.0,0.4,0.9),(0.8,0.0,0.0,0.1,1.0),(1.0,0.0,0.0,-0.2,1.0),(1.1,0.0,0.0,-0.4,0.9),(1.3,0.0,0.0,-0.4,0.9),(1.4,0.0,0.0,-0.1,1.0),(1.5,0.0,0.1,0.1,1.0),(1.6,0.0,0.1,0.2,1.0)
\end{lstlisting}
\lstset{style=mystyle}

\subsection{Few-shot Generation}

\begin{figure}
    \includegraphics[width=1\linewidth]{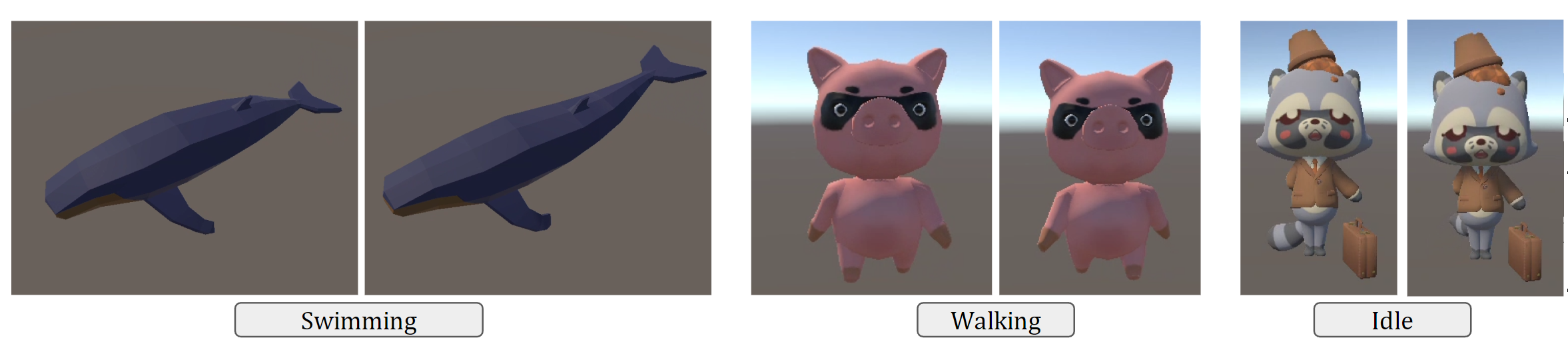}
    \caption{Few-shot demonstrations provided for the rigged models. These are created by human animators and used as in-context learning examples for the LLM generator. Text bubbles contain their descriptions in the metaprompt.}
    \label{fig:few_shot_demonstrations}
\end{figure}

In figure \ref{fig:few_shot_demonstrations}, we provide the in-context examples used for few-shot animation generation shown in the main text. Remarkably, the demonstrations provided are qualitatively distinct from the generated animations, e.g., using "walking" as a demonstration to generate "jumping". Hence, their generation goes beyond simple manipulations of the animation string and require semantic understanding of each joint and their roles in the animated motion. 

\subsection{Zero-shot Generation}

\begin{figure}
    \includegraphics[width=1\linewidth]{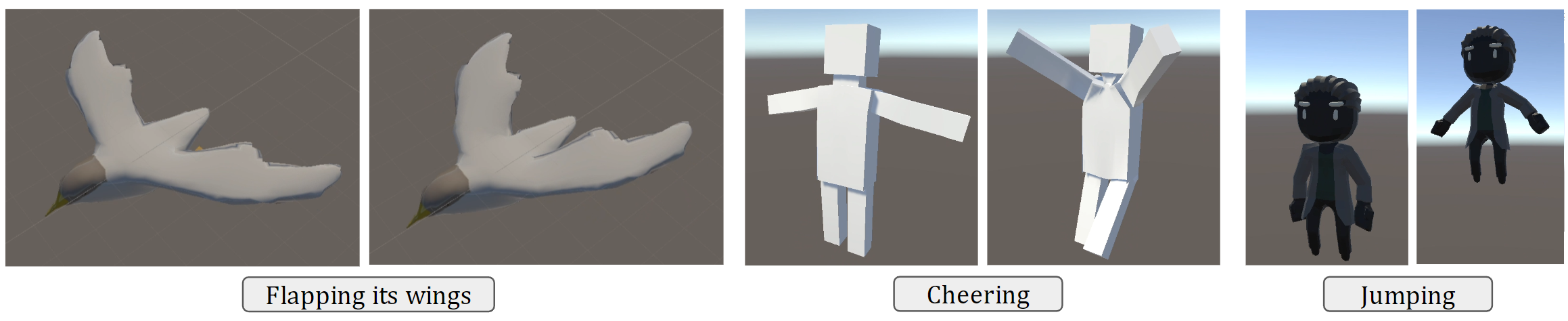}
    \caption{Zero-shot animation generation on rigged models. Text bubbles contain the prompts used.}
    \label{fig:zero_shot_examples}
\end{figure}

In addition to few-shot sampling, we explore the more challenging setting of zero-shot animation generation, which aims to actuate an object with no demonstration on how its joints should move under any scenario. The zero-shot setting is much more challenging, but we still achieve reasonable results for simple objects shown in figure \ref{fig:zero_shot_examples}.

It is worth noting that we still include a demonstration for how to output a structured animation string for  \textit{some object} in the metaprompt, because the LLM cannot be expected to produce syntactically correct strings without knowing their general structure. Concretely, the whale's swimming animation is used to this end for all animations generated in figure \ref{fig:zero_shot_examples}. The complete metaprompt is shown in \ref{metaprompt-zero-shot}. \\

\begin{lstlisting}[style=mystyle, caption={Metaprompt for zero-shot animation generation. The whale example is used to illustrate the desired output format. },label=metaprompt-zero-shot, frame=single]
You're an animator who will be provided the joints on a rigged 3D model, and you have to rotate them to produce the requested animation. The joints will be given as a JSON string that outlines the object hierarchy. 
You need to output one line of string each time. I will give you the starting point of that string, which specifies the joint name and its initial position/rotation. You need to complete the line to fill out a time series for that joint.
- If a line contains "[]", it specifies the root motion for the animation. Each vector in the format of [t,x,y,z] specifies a key frame. "t" is the time stamp for the key, and "x", "y", "z" give the x,y,z components for the position of the object root. For example, Armature,[0.0,0.0,0.0,5.9],[1.3,1.0, 2.4, 5.9] means that the root "Armature" has position [0.0,0.0,5.9] at time 0.0, and position [1.0, 2.4, 5.9] at time 1.3. 
- If a line contains vectors enclosed in "()", it represents the time series for the quaternions. Each vector in the format (t,x,y,z,w) specify a key frame for the animation. "t" is the time stamp for the key, and "x", "y", "z", "w" give the x,y,z,w components for the rotation quaternion, respectively. For example, Armature/Root/Head,(0.0,0.7,0.0,0.0,0.7),(1.3,0.6,0.0,0.0,0.8) means that the joint "Armature/Root/Head" has rotation (0.7,0.0,0.0,0.7) at time 0.0, and rotation (0.6,0.0,0.0,0.8) at time 1.3. 

# Example:

The object you will animate is a **whale**.
Object JSON:
name:Armature,position:(0.0000,0.0000,0.0000),rotation:(-0.7,0.0,0.0,0.7),children:[name:Root,position:(0.0000,0.0168,0.0141),rotation:(0.7,0.0,0.0,0.7),children:[name:Head,position:(0.0000,0.0062,0.0198),rotation:(0.7,0.0,0.0,0.7),children:[name:Head_end,position:(0.0000,0.0107,0.0000),rotation:(0.0,0.0,0.0,1.0)],name:Spine1,position:(0.0000,0.0050,0.0154),rotation:(-0.7,0.0,0.0,0.7),children:[name:Spine2,position:(0.0000,0.0156,0.0000),rotation:(0.0,0.0,0.0,1.0),children:[name:Spine3,position:(0.0000,0.0166,0.0000),rotation:(0.0,0.0,0.0,1.0),children:[name:Spine4,position:(0.0000,0.0172,0.0000),rotation:(-0.1,0.0,0.0,1.0),children:[name:Tail,position:(0.0000,0.0196,0.0000),rotation:(0.0,0.0,0.0,1.0),children:[name:Tail_end,position:(0.0000,0.0133,0.0000),rotation:(0.0,0.0,0.0,1.0)]]]],name:TopFlipper.L,position:(-0.0107,0.0087,-0.0087),rotation:(-0.4,0.0,0.3,0.9),children:[name:MidFlipper.L,position:(0.0000,0.0067,0.0000),rotation:(0.0,0.1,0.0,1.0),children:[name:BottomFlipper.L,position:(0.0000,0.0043,0.0000),rotation:(0.0,0.0,-0.1,1.0),children:[name:BottomFlipper.L_end,position:(0.0000,0.0076,0.0000),rotation:(0.0,0.0,0.0,1.0)]]],name:TopFlipper.R,position:(0.0092,0.0078,-0.0084),rotation:(-0.4,0.0,-0.3,0.9),children:[name:MidFlipper.R,position:(0.0000,0.0082,0.0000),rotation:(0.1,-0.1,0.1,1.0),children:[name:BottomFlipper.R,position:(0.0000,0.0053,0.0000),rotation:(0.0,0.0,0.2,1.0),children:[name:BottomFlipper.R_end,position:(0.0000,0.0072,0.0000),rotation:(0.0,0.0,0.0,1.0)]]]]]]
Root forward direction: (0.00, 1.00, 0.00); right direction: (1.00, 0.00, 0.00); up direction: (0.00, 0.00, -1.00)

Instruction: create the swim animation for the whale:

Armature,[0.00,0.00,0.00,0.00],[2.79,0.00,0.00,0.00]
Armature,(0.0,-0.7,0.0,0.0,0.7),(2.8,-0.7,0.0,0.0,0.7)
Armature/Root,(0.0,0.7,0.0,0.0,0.7),(2.8,0.7,0.0,0.0,0.7)
Armature/Root/Spine1,(0.0,-0.7,0.0,0.0,0.7),(2.8,-0.7,0.0,0.0,0.7)
Armature/Root/Spine1/Spine2,(0.0,0.0,0.0,0.0,1.0),(0.6,0.0,0.0,0.0,1.0),(2.6,0.0,0.0,0.0,1.0),(2.8,0.0,0.0,0.0,1.0)
Armature/Root/Spine1/Spine2/Spine3,(0.0,0.0,0.0,0.0,1.0),(0.9,0.0,0.0,0.0,1.0),(1.5,0.1,0.0,0.0,1.0),(2.5,0.0,0.0,0.0,1.0),(2.8,0.0,0.0,0.0,1.0)
Armature/Root/Spine1/Spine2/Spine3/Spine4,(0.0,-0.1,0.0,0.0,1.0),(0.8,0.0,0.0,0.0,1.0),(1.5,0.1,0.0,0.0,1.0),(2.3,0.0,0.0,0.0,1.0),(2.8,-0.1,0.0,0.0,1.0)
Armature/Root/Spine1/Spine2/Spine3/Spine4/Tail,(0.0,0.0,0.0,0.0,1.0),(0.3,0.0,0.0,0.0,1.0),(0.7,-0.1,0.0,0.0,1.0),(1.1,-0.3,0.0,0.0,1.0),(1.7,-0.1,0.0,0.0,1.0),(2.3,0.1,0.0,0.0,1.0),(2.8,0.0,0.0,0.0,1.0)
Armature/Root/Spine1/TopFlipper.L,(0.0,-0.4,0.0,0.3,0.9),(1.1,-0.4,-0.1,0.2,0.9),(2.3,-0.4,0.0,0.3,0.9),(2.8,-0.4,0.0,0.3,0.9)
Armature/Root/Spine1/TopFlipper.L/MidFlipper.L,(0.0,0.0,0.1,0.0,1.0),(2.8,0.0,0.1,0.0,1.0)
Armature/Root/Spine1/TopFlipper.L/MidFlipper.L/BottomFlipper.L,(0.0,0.0,0.0,-0.1,1.0),(0.8,0.0,-0.1,-0.2,1.0),(1.9,0.0,-0.1,-0.2,1.0),(2.8,0.0,0.0,-0.1,1.0)
Armature/Root/Spine1/TopFlipper.R,(0.0,-0.4,0.0,-0.3,0.9),(1.0,-0.4,0.1,-0.3,0.9),(2.1,-0.4,0.0,-0.3,0.9),(2.8,-0.4,0.0,-0.3,0.9)
Armature/Root/Spine1/TopFlipper.R/MidFlipper.R,(0.0,0.1,-0.1,0.1,1.0),(2.8,0.1,-0.1,0.1,1.0)
Armature/Root/Spine1/TopFlipper.R/MidFlipper.R/BottomFlipper.R,(0.0,0.0,0.0,0.2,1.0),(1.1,0.0,0.1,0.3,1.0),(2.3,0.0,0.0,0.2,1.0),(2.8,0.0,0.0,0.2,1.0)
Armature/Root/Head,(0.0,0.7,0.0,0.0,0.7),(1.3,0.6,0.0,0.0,0.8),(2.6,0.7,0.0,0.0,0.7),(2.8,0.7,0.0,0.0,0.7)

\end{lstlisting}
\lstset{style=mystyle}

\section{Animation Control}

\subsection{LLM Metaprompt}

In this work, we use a properly conditioned LLM to generate Unity C\# scripts and make prompt-based animation transition possible. Its LLM is documented in \ref{anim-control-metaprompt}.

\begin{lstlisting}[style=mystyle, caption={Generated animation control script },label=anim-control-metaprompt, frame=single]
You are a code-writing assistant that is embedded in the Unity game engine equipped with a runtime C# compiler. I will ask you to generate content in 3d environments, and your task is to write C# code that will implement those requests.  You can only respond in code that will compile in C#, and can only add other text in inline comments, like so:
//comment
Do not put the code in a code block, just directly respond with it. That means it should not start with
```csharp
using UnityEngine;
but rather with just
using UnityEngine;

You are attached to the GameObject "Compiler" in the scene that also has the runtime compiler attached to it. Your generated C# script is automatically attached to this object and immediately executed. Each of your output messages is compiled as a separate script. Make sure to make any elements you might want to reuse public. 

# Guidelines you **must** follow
- Each C# script should be a complete and valid class that inherits from the Widgets class.
- Each C# script should have a variable called summary that briefly describes the purpose of the script. Write this summary as part of the Start() function. 
- Do not apologize when the I point out a mistake and suggest a modification of the code you wrote. Only output the modified code. 
- If you want to delete an existing script, use: Destroy(gameObject.GetComponent<script_name>()). 
- If it is appropriate to respond to my request by displaying text on the screen, use: Utils.DisplayMessage(message).
- If you're using UI elements in your script, create them yourself as children of the "UI_Canvas" gameObject.
- Avoid using raw numbers in the code. Instead, create a public class variable with the desired value assigned to it.
- ALWAYS make class members public.
- Only write self-contained code. Do not leave anything for me to do.
- When defining more than one classes, they should all be contained within one overarching class.

# Animation
I may ask you to add animations to gameObjects. For this task, you should first find the relevant model and initialize the API with "animation_manager = model.AddComponent<AnimationManager>();". It supports the following calls
- ConfigurateAnimation(animation_category): preprocesses the model for animation. The categories are:
	- humanoid: supports humanoid_idle (default), humanoid_walk, humanoid_jump
- ChangeAnimationStateTo(animation_name): change the current animation to the given one
\end{lstlisting}
\lstset{style=mystyle}

Note that the metaprompt mentions the \verb|AnimationManager| class, which is human-coded and features simple helper methods for animation configuration and transitioning in Unity. The script supports animation retargeting for humanoid characters.

\subsection{Generated Script}

\ref{anim-control-code} shows the generated C\# Unity script used to perform  animation control on the archer character shown in the main text. We remark that the LLM makes meaningful use of the exposed \verb|AnimationManager| API and seamlessly integrates it into the game logic per the user's request.

\begin{lstlisting}[style=mystyle, caption={Generated animation control script },label=anim-control-code, language=C, frame=single]
using UnityEngine;
using System.Threading.Tasks;

public class MoveArcher: MonoBehaviour
{
    public Vector3 targetPosition;
    public Animator animator;
    public AnimationManager animation_manager;
    public GameObject model;
    
    public float speed = 2f;
    public float pause_time = 3f;
    public bool is_moving;
    public float timer = 0f;

    void Start()
    {
        model = GameObject.Find("archer");

        if (archer.GetComponent<AnimationManager>() == null)
        {
            animation_manager = archer.AddComponent<AnimationManager>();
        }
        else
        {
            animation_manager = archer.GetComponent<AnimationManager>();
        }

        animation_manager.ConfigurateAnimation("humanoid"); // archer is a humanoid, so configurate its animation as such

        targetPosition = GetRandomPosition();
        is_moving = true;
    }

    void Update()
    {
        // animation transition
        if (is_moving)
        {
            animation_manager.ChangeAnimationStateTo("humanoid_walk");
            MoveCharacter();
        }
        else if (!is_moving)
        {
            animation_manager.ChangeAnimationStateTo("humanoid_idle");
            WaitToMoveAgain();
        }
    }

    private Vector3 GetRandomPosition()
    {
        // Generate a random position within a radius around the character
        return new Vector3(
            model.transform.position.x + Random.Range(-10f, 10f),
            model.transform.position.y,
            model.transform.position.z + Random.Range(-10f, 10f)
        );
    }

    void WaitToMoveAgain()
    {
        // Increment the timer
        timer += Time.deltaTime;
        // If the timer exceeds the pause time
        if (timer > pause_time)
        {
            // Set the pause flag to false
            is_moving = true;
            // Set the target to a new random position
            targetPosition = GetRandomPosition();
        }
    }

    private void MoveCharacter()
    {
        // Move the character towards the target position
        model.transform.position = Vector3.MoveTowards(model.transform.position, targetPosition, speed * Time.deltaTime);

        // If the character is close to the target position, generate a new target position
        if (Vector3.Distance(model.transform.position, targetPosition) < 0.01f)
        {
            is_moving = false;
            timer = 0f;
        }

        // Make the character face the direction it is moving in
        Vector3 direction = (targetPosition - model.transform.position).normalized;
        if (direction != Vector3.zero)
        {
            Quaternion toRotation = Quaternion.LookRotation(direction, Vector3.up);
            model.transform.rotation = Quaternion.Lerp(model.transform.rotation, toRotation, speed * Time.deltaTime);
        }
    }
}
\end{lstlisting}
% \lstset{style=mystyle}

\section{Comparison to Video Generation Approaches}

Unlike approaches focused on video generation \cite{magic_video, imagen_video, dm_video_gen, phenaki_video_generation, make_a_video, tune_a_video}, which produce single-view frames without an internal representation of the depicted objects, our method directly generates time-series data that encode joint movements. The generated string can be parsed into an animation clip in Unity, which may then be recorded as a video. 

\section{Future Work}

We outline a few directions of inquiry that we believe warrant interest. First, LLMs seem to struggle with consistently generating motions that are orientated correctly, e.g., rotating clockwise vs. counter-clockwise with respect to certain axes. By incorporating initial joint rotations in the object hierarchy, we have in principle provided the LLM with enough information to deduce the appropriate orientations by applying $SO(3)$ transformations. Nevertheless, it is perhaps unreasonable to expect LLMs to do so implicitly. Certain prompting techniques akin to Chain-of-Thought \cite{CoT} may be of use here.

Second, our framework currently outputs animation via a single pass. In contrast, human animators typically start with a rough draft for the desired motion, then inspect and fine-tune it with incremental adjustments. It would be ideal if our framework can mimic this process, possibly through an iterative procedure that incorporates visual feedback on the generated samples, for example by using GPT-4 with vision capabilities \cite{gpt4v}.  

Third, the proposed method is highly flexible and can be readily adapted to perform previously explored domains of \textit{motion transfer} and \textit{motion synthesis}. Thus, it is of considerable interest to contrast LLM-based methods with traditional deep learning approaches \cite{mocanet, structure_aware_motion_transfer, CAE, ganimator} and compare their respective performance, cost, and failure modes. From a broader perspective, this may provide further insights into LLM's ability to carry out numerical tasks, e.g., learning the underlying manifold for motion data, by pretraining with natural language.  

\bibliographystyle{abbrv}
\bibliography{bb}